# Carrier scattering, mobilities and electrostatic potential in mono-, bi- and tri-layer graphenes

Wenjuan Zhu*, Vasili Perebeinos, Marcus Freitag and Phaedon Avouris**
IBM Thomas J. Watson Research Center, Yorktown Heights, NY 10598, USA

## ABSTRACT

The carrier density and temperature dependence of the Hall mobility in mono-, bi- and tri-layer graphene has been systematically studied. We found that as the carrier density increases, the mobility decreases for mono-layer graphene, while it increases for bi-layer/tri-layer graphene. This can be explained by the different density of states in mono-layer and bi-layer/tri-layer graphenes. In mono-layer, the mobility also decreases with increasing temperature primarily due to surface polar substrate phonon scattering. In bi-layer/tri-layer graphene, on the other hand, the mobility increases with temperature because the field of the substrate surface phonons is effectively screened by the additional graphene layer(s) and the mobility is dominated by Coulomb scattering.

We also find that the temperature dependence of the Hall coefficient in mono-, bi- and tri-layer graphene can be explained by the formation of electron and hole puddles in graphene. This model also explains the temperature dependence of the minimum conductance of mono-, bi- and tri-layer graphene. The electrostatic potential variations across the different graphene samples are extracted.

## I. INTRODUCTION

In the past few decades, the semiconductor industry has grown rapidly by offering every year higher function per cost. The major driving force of this performance increase is device scaling. However, scaling is becoming more and more difficult and costly as it approaches its scientific and technological limits. An alternative path for future development is needed. Some innovations use new computational state variables, such as spins or magnetic flux instead of charges [ref. 1, 2]. Other approaches adopt new materials such as carbon nanotubes and graphene to replace silicon [ref. 3]. In 2004, single atomic layer graphene was first produced by mechanical exfoliation [ref. 4 ]. This enabled researchers to access and to study this promising material.

Graphene is a 2D material containing carbon atoms tightly bonded together in a honeycomb arrangement [ref. 3]. The dispersion relation for mono-layer graphene is linear, $E_F = \hbar v_F k$, with a Fermi velocity $V_F = 10^6\ m/s$. Mono-layer graphene has zero band gap and zero effective mass. Bi-layer graphene, on the other hand, has a parabolic band structure with an effective mass $m^*_{BG}=0.037m_0$ [ref. 5 ], determined by the inter-layer coupling. The band gap in bi-layer graphene can be varied by means of an external perpendicular field [ref. 6, 7, 8]. Tri-layer graphene has a similar dispersion relation as bi-layer graphene with parabolic bands, except that the effective mass is larger $m^*_{TL}=0.052m_0$ [ref. 5]. Recently it was proposed that tri-layer graphene is semi-metallic with a tunable band overlap [ref. 9].

* wenjuan@us.ibm.com
** avouris@us.ibm.com

The key property of interest in graphene for electronic applications is the fast electronic transport expressed by its high carrier mobility. Suspended mono-layer graphene has been shown to have extremely high mobilities (up to 200,000 $cm^2/Vs$) [ref.10], however, this value is strongly reduced in the supported structure by impurity and phonon scattering [ref.11,12,13,14]. Since mono-layer graphene has no band-gap, it is not directly suitable for digital electronics, but is very promising for analog, high frequency (RF) applications [ref.15] and interconnects [ref.16, 17]. Transport in bi-layer and trilayer graphene has been studied less extensively. Because of their different band-structures and screening properties, the contributions of the various scattering mechanisms are expected to change in these layers and different electronic applications are possible. For example, bi-layer graphene could at high fields develop a significant band-gap to be employed in digital electronics [ref.8].

Here we present a systematic study of the Hall carrier mobilities and their temperature dependence for mono-, bi- and tri-layer graphenes in order to determine the importance of the different scattering mechanisms in limiting these mobilities at technologically relevant carrier densities. We also use Hall-effect measurements to determine the electrostatic potential variations in the graphene layers at low carrier densities.

## II. EXPERIMENT

The graphene layers were deposited through mechanical exfoliation of graphite on a 300nm $SiO_2$ film grown on a silicon substrate. The number of layers deposited was determined by the changes in the reflectance of green light [ref. 9, 18] and by Raman spectroscopy (see Appendix). The Hall-bar geometry was fabricated using oxygen plasma, while the electrodes were made of Ti/Pd/Au. The Si-substrate itself was used as a back gate. The magnetic field was +/- 2Tesla and the samples were measured in high vacuum in the temperature range of 4.2K to 350K. The carrier density $n_s$ was extracted from the Hall voltage $V_H$: $n_s = IB/q|V_H|$, where I is current, B is the magnetic field, and q is the electron charge. The Hall mobility was extracted using the relation: $\mu = \sigma_{xx}/qn_s$, where $\sigma_{xx}$ is the four-probe conductivity along the current direction.

## III. MOBILITY AND CARRIER SCATTERING

The mobility and scattering mechanism in mono-layer graphene has been studied both experimentally [ref. 11, 19,20] and theoretically considering Coulomb scattering [ref.21,22], short-range scattering [ref.12], phonon scattering by graphene phonons [ref. 13, 23], substrate surface phonon scattering [ref. 24], mid-gap states[ref.25], and roughness [ref.14]. There are some reports on mobility extracted from two terminal measurements on bi-layer [ref. 26] and tri-layer graphene [ref.9]. However, the temperature dependence of Hall mobility and the scattering mechanisms for bi-layer and tri-layer graphene has not been systematically established yet. In this section, we will discuss the temperature dependence of Hall mobility for bi-layer and tri-layer graphene and compare it with mono-layer graphene.

Figure 1 shows the carrier density dependence of the mobility at various temperatures (from 4.2K to 350K) for mono-layer, bi-layer and tri-layer graphene, respectively. We see that, as the carrier density increases, the mobility decreases for mono-layer graphene, while it increases for bi-layer and tri-layer graphene.

The temperature dependence of the hole mobility at various carrier densities for these layers is shown in Figures 2a,b,c. We observe that the mobility decreases with temperature for mono-layer graphene, especially when the temperature is above ~200K, while it increases with temperature for bi-layer and tri-layer graphenes.

These different trends of the carrier mobility in mono-layer and bi-layer/tri-layer graphenes can be understood by differences in their density-of-states and the additional screening of the field of substrate surface phonons in bi-layer/tri-layer graphenes. A detailed analysis is given below.

## A. Carrier density dependence of the mobility at low temperatures

At low temperatures, the dominant scattering mechanism is impurity scattering which includes Coulomb scattering and short-range scattering $\mu_{total}{}^{-1} \approx \mu_c{}^{-1} + \mu_{sr}{}^{-1}$ [ref. 12]. The scattering time for Coulomb and short-range scattering can be expressed as [ref. 12]:

$$\frac{1}{\tau(\varepsilon_k)} = \frac{\pi}{\hbar}\sum_{k'} n_i \left|\frac{V(q)}{\varepsilon(q)}\right|^2 (1-\cos^2\theta)\delta(\varepsilon_k - \varepsilon_{k'}) \qquad (1)$$

where $q = |k - k'|$, and θ is the angle between k and k', $V^a(q)$ is the matrix element of the scattering potential between an electron and an impurity, $\varepsilon(q)$ is the 2D finite temperature static RPA dielectric (screening) function appropriate for graphene, $n_i$ is the concentration of the a-th kind of impurity center. If we assume the screening function ε is q-independent or very weakly q-dependent, then we have $1/\tau(\varepsilon_k) \propto V(k)D(\varepsilon_k)$. The energy average scattering time can be written as:

$$\langle\tau\rangle = \int d\varepsilon_k \varepsilon_k \tau(\varepsilon_k)(-\frac{\partial f}{\partial \varepsilon_k}) / \int d\varepsilon_k \varepsilon_k (-\frac{\partial f}{\partial \varepsilon_k}) \text{ [ref. 12]}. \qquad (2)$$

At low temperatures, the energy averaging of the scattering time is $\langle\tau\rangle \approx \tau(\varepsilon_F)$ .The matrix element is given by $V(q,d) = \frac{2\pi e^2}{\kappa q}$ for Coulomb scattering, if we assume the scattering charge centers are at the $SiO_2$-graphene interface, while it is constant for short-range scattering [ref. 12].

The density-of-states ( $D(E_F)$ ) in mono-layer graphene is proportional to $E_F$ [ref.22]:

$$D(E_F) = \frac{2E_F}{\pi(\hbar V_F)^2} \tag{3}$$

while in bi-layer or tri-layer graphene it is constant: [ref.22]

$$D(E_F) = \frac{2m^*}{\pi\hbar^2} \tag{4}$$

Here the Fermi energy $E_F \propto k \propto \sqrt{n}$ for mono-layer [ref. 12], while $E_F \propto k^2 \propto n$ for bi-layer or tri-layer graphene. Note that in the Boltzmann treatment the mobility is related to the scattering time as $\mu = \frac{eD(E_F)v_F^2\langle\tau\rangle}{2n}$, which reduces to $\mu_{SL} = \frac{ev_F^2\langle\tau\rangle}{E_F}$ in a mono-layer and to $\mu_{ML} = \frac{e\langle\tau\rangle}{m^*}$ in bi-layer and tri-layer graphene.

For mono-layer graphene, the mobility limited by Coulomb scattering was found to be independent of carrier density, $\mu_{c_SL} \propto \text{constant}$, and the mobility limited by short-range scattering was found to be inversely proportional to the carrier density for mono-layer, $\mu_{sr_SL} \propto 1/n$ [ref.12].

For bi- and tri-layer graphene, based on equation (1,2,4), we find that the mobility limited by Coulomb scattering is proportional to the carrier density, $\mu_{c_ML} \propto n$, while the mobility limited by short-range scattering is constant: $\mu_{sr_ML} \propto \text{constant}$. These considerations explain why the mobility at 4.2K for mono-layer graphene decreases with increasing carrier density, while it increases with increasing carrier density for bi- and tri-layer graphene, as shown in Figures 1 (a)-(c).

## B. Temperature dependence of mobility

The mobility in mono-layer graphene decreases rapidly with increasing temperature when the temperature is above about 200K (see Figure 2). This is primarily due to scattering by thermally-excited surface phonons of the $SiO_2$ substrate [ref. 11, 24, 27]. The $SiO_2$ optical phonons at the substrate/graphene interface modulate the polarizability which produces an electric field that couples to the carriers in graphene. The coupling or the field depends exponentially on the substrate graphene distance. At the Van der Waals distance of about 3.5 A it is much stronger than the coupling of the carriers to the acoustic phonons of graphene. There are two important surface phonons in $SiO_2$ with energies of about 59meV and 155meV [ref.24,28]] and the coupling is determined by the dielectric polarization field: $\vec{P} \propto \sqrt{\hbar\omega_{so}\left(\frac{1}{\varepsilon_\infty+1} - \frac{1}{\varepsilon_0+1}\right)}$ where $\omega_{so}$ is a surface optical phonon frequency, $\varepsilon_0$ and $\varepsilon_\infty$ are the low- and high-frequency dielectric constants of $SiO_2$ correspondingly, and the dielectric constant of air is one. The surface polar phonon scattering is proportional to the phonon population number such that the scattering time can be expressed as: $\tau_{ox}^{-1} \propto \sum_i \frac{c_i}{e^{\hbar\omega_i/\kappa_B T}-1}$ , where for $SiO_2$ the ratio of $c_2/c_1 \approx 6.5$ [ref.24] is determined by the dielectric constants in $SiO_2$. Thus, as the temperature is increased, the mobility is expected to decrease drastically. For bi- and tri-layer graphene, however, the mobility increases instead of decreasing as the temperature increases. This is due to the fact that the substrate surface phonon induced field is effectively screened by the additional graphene layer(s). This makes bi- and tri-layer graphenes very promising as high mobility materials for electronics.

In bi- and tri-layer graphene samples, the temperature dependence of mobility is mainly determined by Coulomb scattering. Due to the parabolic band structure, the energy averaging of the Coulomb scattering time can result in the mobility increasing proportionally to temperature: $\mu \propto k_B T$ [ref.29]. The dielectric screening, which we ignored in the above analysis, could also introduce an additional temperature dependence. For mono-layer, however, it was found that the temperature dependence of Coulomb scattering is very weak, when $k_B T << E_F$ [ref. 21], which is the temperature and carrier density range we are investigating here.

The temperature dependence of the mobility limited by short-range scattering is independent of temperature for bi-layer and tri-layer graphene, since the density-of-states, the matrix element and the screening function are all energy independent. In mono-layer graphene, the temperature dependence of conductivity or mobility that is limited by short-range scattering is nearly constant, when $k_B T << E_F$ [ref.22]. On the other hand, the mobility limited by the graphene acoustic phonons in mono-layer graphene is inversely proportional to temperature [ref.30]. However, since the magnitude of the mobility limited by graphene phonon scattering is of the order of $10^5$ $cm^2/V\text{-}s$ [ref. 13], i.e. much larger than the mobilities limited by the other three scattering mechanisms discussed above, it can be neglected.

Based on the above discussion, we fit the measured carrier mobilities using the following model for mono-layer and bi- or tri-layer graphene.

For mono-layer graphene, at 4.2K, the mobility can be expressed as: $\mu_{4.2K}^{-1} \approx \mu_c^{-1} + \mu_{sr}^{-1}$, where $\mu_c = S_c$ and $\mu_{sr} = \frac{S_{sr}}{n}$ with $S_c$, $S_{sr}$, as fitting parameters determined by the Coulomb and short-range scattering respectively. At high temperatures, the mobility can be expressed as:

$$\mu^{-1} \approx \mu_{4.2K}^{-1} + \mu_{gr}^{-1} + \mu_{ox}^{-1} \qquad (5)$$

where $\mu_{gr} = \frac{S_{gr}}{n*T}$ and $\mu_{ox} = S_{ox} n^{\alpha} \left( \frac{1}{e^{(59meV)/\kappa_B T}-1} + \frac{6.5}{e^{(155meV)/\kappa_B T}-1} \right)^{-1}$ with $S_{gr}$ , $S_{ox}$ , and $\alpha$ as fitting parameters determined by the graphene acoustic phonon and substrate surface polar phonon scattering respectively.

For bi- and tri-layer graphene, the mobility can be expressed as:

$$\mu^{-1} \approx \mu_c^{-1} + \mu_{sr}^{-1} \quad (6)$$

where we find $\mu_c = (A + B \cdot T) \cdot n$ and $\mu_{sr} = C$ with A, B and C are fitting parameters.

The fitting results for mono-, bi- and tri-layer graphenes are shown in Figures 1 (a)-(c) and 2 (a)-(c). The symbols are the measured data and the lines are the fits. We see that these formulae fit the measured data very well.

One important point is that the mobility limited by Coulomb and short-range scattering for bi-layer/tri-layers is inversely proportional to the square of the effective mass, $(m^*)^2$. This means the more graphene layers, the heavier the effective mass, leading to a higher degradation of the mobility limited by Coulomb and short-range scattering for the same impurity concentration. This is currently a disadvantage for multi-layer graphenes. However, when the impurity concentration of the samples and the substrate are significantly reduced by process optimization and since the surface phonon scattering in bi- or tri-layer graphene is largely screened, the mobility for bi- and tri-layer graphene can be significantly higher at room temperature in unsuspended devices. Note that graphite has the highest mobility reported so far [ref.31], although it can not be switched off. In addition, electrical noise, which is very important in electronic applications, is significantly reduced in bi- and tri-layers [ref.32].

## IV. HALL COEFFICIENT

Besides the Hall mobility, another important aspect of the current transport is the carrier density $n_s$, which is determined by the electronic structure and can be extracted from the Hall coefficient $R_H$ ($n_s = \frac{1}{qR_H}$, when only one type of carrier is dominant). The Hall coefficient is defined as $R_H = V_H / I_H B$, where $V_H$ is the measured Hall voltage, $I_H$ is the constant current source and B is the applied magnetic field. To investigate the temperature dependence of carrier density, the temperature dependence of Hall coefficient is analyzed.

Figure 3 (a) to (c) shows the Hall coefficient as a function of the back-gate voltage, $V_{BG}$-$V_{Dirac}$, at various temperatures (from 4.2K to 350K) for mono-layer, bi-layer and tri-layer graphene, respectively. As the temperature is increased, the height of the peak is reduced (or the slope of $|R_H|$ vs. Vg decreases) for all graphene layers.

This could be explained by either band overlap or by the formation of electron and hole puddles in graphene near the Dirac point. However, no band overlap is expected in mono-layer and in bi-layer graphene, therefore we suggest that puddle formation is the dominant mechanism for the observed temperature dependence. The electron and hole puddles were previously observed in scanning tunneling spectroscopy measurements [ref. 33, 34]. The formation of electron and hole puddles was attributed to the intrinsic ripples [ref. 16, 35, 36] in graphene and extrinsic charge-induced inhomogeneites in the carrier density[ref.33, 37].

The ambipolar Hall coefficient is given by [ref. 38]:

$$qR_H = \frac{\mu_e^2 n_e - \mu_h^2 n_h}{\left(\mu_e n_e + \mu_h n_h\right)^2} \tag{7}$$

The gate voltage is related to the carrier density by following equation [ref.39]:

$$V_g = q(n_e - n_h)(\frac{1}{C_{ox}} + \frac{1}{C_q}) \tag{8}$$

where $C_{ox}$ is the oxide capacitance and $C_q$ is the quantum capacitance. For mono-layer graphene, the quantum capacitance is $C_q = \frac{2Ee^2}{\pi\hbar^2 v_F^2}$, where $v_F$ is the Fermi velocity. For bi-layer and tri-layer graphene, $C_q = \frac{2m^* e^2}{\pi\hbar^2}$, where $m^*$ is the effective mass. For a 300 nm $SiO_2$, the quantum capacitance $C_q >> C_{ox}$ when carrier density is larger than $2x10^{11}cm^{-2}$ for mono-layer graphene and $C_q >> C_{ox}$ at any carrier density for bi- and tri-layer. Therefore, equation (8) can be reduced to $q(n_e\text{-}n_h)=C_{ox}V_g$. For large gate voltages, when only one type of carriers is present, Eq.(7) reduces to $R_H = \frac{-1}{C_{ox}V_g}$, which is used to determine the gate oxide capacitance for the electron and hole branches. In the vicinity of the Dirac (mid-gap) point, which we define as the point where $R_H=0$, we assume that the hole and electron carrier density are equal to each other, i.e. $n_e=n_h=n_{Dirac}/2$, where $n_{Dirac}$ is the total carrier density at the Dirac point.

For mono-layer graphene, at low carrier density, since the dominant scattering is Coulomb scattering $\mu_c$ , which is independent of carrier density and the measured electron and hole mobilities are roughly equal to those at high carrier density, we can assume $\mu_e=\mu_h$ near Dirac point, and Eq. (7) reduces to:

$$qR_H \approx \frac{C_{ox}V_g}{qn_{dirac}^2} \tag{9}$$

For bi-layer and tri-layer graphene, the mobility limited by Coulomb scattering is proportional to carrier density ($\mu_c \propto n$) and the total mobility $\mu_{total} \approx \mu_c$, therefore equation (7) is reduced to:

$$qR_H \approx \frac{3C_{ox}V_g}{qn_{Dirac}{}^2} \tag{10}$$

If we assume that the area of the hole and electron puddles is equal in size and simplify the spatial electrostatic potential to a step function with the characteristic peak to peak height of +/-$\Delta$, as illustrated in figure 4, then the total carrier density at the Dirac point is:

$$n_{dirac} = 2n_e = 2n_h = \int_{-\Delta}^{\infty} D(E+\Delta)\frac{1}{e^{E/k_BT}+1}dE + \int_{\Delta}^{\infty} D(E-\Delta)\frac{1}{e^{E/k_BT}+1}dE \quad (11)$$

For mono-layer graphene, in the limit Δ/kT>>1 equation (11) can be simplified as:

$$n_{dirac} \approx \frac{2}{\pi\hbar^2 v_F^2}\left(\frac{\Delta^2}{2} + \frac{\pi^2}{6}k_B^2T^2\right) \quad (12)$$

For bi-layer and tri-layer, from equation (4) and (11), we obtain:

$$n_{dirac} = \frac{2m^*}{\pi\hbar^2}k_BT[\ln(1+\exp(\Delta/k_BT)+\ln(1+\exp(-\Delta/k_BT)] \quad (13)$$

We can extract the electrostatic potential Δ using equation (9) and (12) for mono-layer graphene and equation (10) and (13) for bi- and tri-layer graphene. Figure 5 shows the fitting for mono-, bi- and tri- layer graphene. From the fitting, we get $v_F$=1.3 x $10^6$ m/s and Δ=54 meV for mono-layer graphene, $m^*$=0.063, Δ =31 meV for bi-layer graphene and $m^*$=0.082 and Δ =43 meV for tri-layer graphene. These results are consistent with the scanning tunneling spectroscopy measurements on mono-layer graphene [ref. 33], which reported a maximum variation of the Dirac point by $\Delta_{max}$ = 77 meV.

Note that another possible cause of the temperature dependence of the Hall coefficient is band overlap. It was proposed that the bi-layer graphene is a semi-metal with band overlap based on the $R_H$ vs $V_g$ results and temperature dependence of minimum conductance [ref.9]. However, if we use the band overlap model, we will obtain a band-overlap larger than 40meV in all three cases: mono-layer, bi-layer and tri-layer graphene. Note that we only applied a back-gate voltage and the data region used for fitting is the region near the Dirac point (in-between two $R_H$ peaks, i.e. the mixed carrier region only). The perpendicular field is nearly zero. At this field, both mono-layer and bi-layer graphenes are known to be zero band-gap materials [ref.8]. Therefore, it is very unlikely that this temperature dependence of the Hall coefficient is due to band overlap. We think this temperature dependence of the Hall coefficient near the Dirac point is most likely due to the electron-hole puddles formation. This finding cast doubt on the proposed value of the band overlap in tri-layer graphene [ref. 9]. It is clear that local probes like STS are needed to resolve this issue.

It should also be mentioned that the carrier density at Dirac point $n_{Dirac}$ increases with increasing number of layers at a given temperature, as revealed in Figure 5. This is because the density-of-states near the Dirac point increases with increasing number of layers, as illustrated in the inset of Figure 5. Since the "on" state carrier density $n_{on} \approx (V_g-V_{Dirac})C_{ox}/q$ is independent of the number of graphene layers, the carrier density on/off ratio ($n_{on}/n_{Dirac}$) will decrease with the number of graphene layers. This is one important factor driving the reduction of the current on/off ratio ($\frac{I_{on}}{I_{off}} = \frac{n_{on}}{n_{Dirac}}\frac{\mu_{on}}{\mu_{Dirac}}$) as the number of layers of graphene increases and restricts the upper limit of the maximum number of graphene layers to be used for a given on/off ratio requirement.

# V. MINIMUM CONDUCTANCE

Based on the temperature dependence of mobility and carrier density discussed in the previous two sections, now we can use those models to explain the temperature dependence of minimum conductance.

The minimum conductances as a function of temperature for mono-layer, bi- and tri-layer graphene are shown in the Figure 6. As the temperature increases, the minimum conductance increases dramatically for bi-layer and tri-layer graphene, while it is nearly unchanged for mono-layer graphene. This can be explained by the temperature dependence of the carrier density and temperature dependence of the mobility.

For mono-layer graphene, the carrier density at the Dirac point is proportional to $T^2$ according to equation (12), while the mobility decreases with temperature due to phonon scattering. As a result, the conductance $\sigma = en\mu$ is very weakly temperature dependent.

For bi-layer and tri-layer graphene, however, both the carrier density and the mobility increase with temperature, as shown in equation (13) and (6). Assuming that $\mu_e = n_e(A_e + B_e \cdot T)$ and $\mu_h = n_h(A_h + B_h \cdot T)$, the conductance can be written in the form:

$$\sigma(T) = P * T^2 [\ln(1 + e^{\Delta/(k_B T)}) + \ln(1 + e^{-\Delta/(k_B T)})]^2 (1 + r * T) \qquad (14)$$

where $\Delta$ and P are fitting parameters. The parameter $r = (B_e + B_h)/(A_e + A_h)$, where A and B were extracted previously from the Hall mobility fitting, assuming the temperature dependence of mobility at the Dirac point is similar to the one at higher carrier density. The insets in Figures 6 (b) and (c) show the measured minimum conductance as a function of temperature and fitting results. The red lines are fits considering the temperature of carrier density only, while the green lines are fits considering the temperature dependence of both carrier density and mobility. As we can see, the fitting is better when we consider the temperature dependence of both carrier density and mobility. From these fits (including the density dependence of mobility) we obtain $\Delta$=26meV and $m^*$=0.045 for bi-layer and $\Delta$=36meV and $m^*$=0.066 for tri-layer graphene. These results are consistent with the results extracted above from the $R_H$ fitting.

# VI. SUMMARY

We have performed systematic studies of the transport properties and carrier scattering mechanisms in mono-, bi- and tri-layer graphenes as a function of carrier density and temperature using Hall-effect measurements. We found that as the carrier density increases, the mobility decreases for mono-layer graphene, while it increases for bi-layer and tri-layer graphene. This can be ascribed to the different density-of-states for mono-layer and bi-layer/tri-layer graphenes. As the temperature increases, we find that the mobility decreases for mono-layer graphene due to the surface polar substrate scattering as in ref. [11], while it increases almost linearly with temperature (see Eq. (6)) for bi-layer/tri-layer graphene. This is attributed to the fact that Coulomb scattering decreases with temperature for bi-layer/tri-layer graphene due to their parabolic band structure and screening. Furthermore, scattering by the $SiO_2$ polar substrate surface phonons is effectively screened in bi-layer/tri-layer. We also found that the temperature dependence of the Hall coefficient in mono-, bi- and tri-layer graphene can be explained by the formation of electron and hole puddles in the graphene. This model also explains the temperature dependence of the minimum conductance of mono-, bi- and tri-layer graphene. The variation of the electrostatic potential along the surface and the effective masses for bi- and tri-layer graphenes are extracted from these measurements.

## ACKNOWLEDGEMENT

We gratefully acknowledge F. Xia, Y.-M. Lin, D. B. Farmer, H.-Y. Chiu for insightful discussions. We would also like to thank B. Ek and J. Bucchignano for their help with device fabrication.

## APPENDIX

The number of layers in the graphene samples was determined by the green light reflectance shift and the Raman spectrum. The green light reflectance shift method is based on the optical contrast between graphene and the $SiO_2$ substrate. This has been demonstrated to be an efficient and reliable method to determine the number of graphene layers [ref.18, 9]. The green light reflectance shift is defined as $GRS=(G_s-G_f)/G_s$, where $G_f$ is the green-channel component of the RGB value on the flake and $G_s$ is the corresponding value on the substrate. Figure 7 (a) shows the green light reflectance shift for mono-layer, bi-layer and tri-layer graphene. The average shifts for mono-layer, bi-layer and tri-layer graphene are $GRS_{MG}=0.053$, $GRS_{BG}=0.097$, $GRS_{TG}=0.134$, respectively. The red, green and blue dots are the flakes measured by Raman scattering as well. The Raman spectrum is shown in Figure 7 (b). While it is difficult to differentiate between bi-layer and tri-layer graphene from the Raman spectra alone, the mono-layer graphene stands out with a large G-prime to G ratio in agreement with the green light shift method. [ref. 40]

## FIGURE CAPTION

FIG. 1: Hall mobility as a function of carrier density at temperatures from 4.2K to 350K in (a) mono-layer graphene, (b) bi-layer graphene and (c) tri-layer graphene. The symbols are the measured data, the lines are fits.

FIG. 2: Hall mobility for holes as a function of temperature at various carrier densities in (a) mono-layer graphene, (b) bi-layer graphene and (c) tri-layer graphene. The symbols are the measured data, the lines are fits.

FIG. 3: Hall coefficient as a function of $V_{BG}$-$V_{Dirac}$ in (a) mono-layer graphene, (b) bi-layer graphene and (c) tri-layer graphene.

FIG. 4: Illustration of the spatial inhomogeneity of the electrostatic potential and the model used in the analysis of the potential variation for mono-layer, bi-layer and tri-layer graphene.

FIG. 5: Carrier density at the Dirac point extracted from the Hall coefficient as a function of temperature in single-, bi- and tri-layer graphene. The symbols are the measured data, the lines are fits using equation (12) for mono-layer graphene and equation (13) for bi-layer and tri-layer graphene.The inset illustrates the density-of-states for single-, bi- and tri-layer graphene.

FIG. 6: Minimum conductance as a function of temperature for (a) mono-layer, (b) bi-layer and (c) tri-layer graphene. The symbols are the measured data. In (b) and (c), the red lines are fits that consider only the temperature dependence of the carrier density $n_{Dirac}(T)$ using equation (13), while the green line fits takes into account the temperature dependence of both carrier density and mobility $n_{Dirac}(T)\mu(T)$ using equation (14).

FIG. 7: (a) Green light reflectance shift, (b) Raman spectrum of mono-layer, bi-layer, tri-layer graphenes. The red, green and blue dots in (a) are the flakes measured with Raman as shown in (b). The Raman spectra are offset for clarity. The inset in (b) shows the zoomed-in spectrum of the G-prime band, scaled to the maximum intensity value.

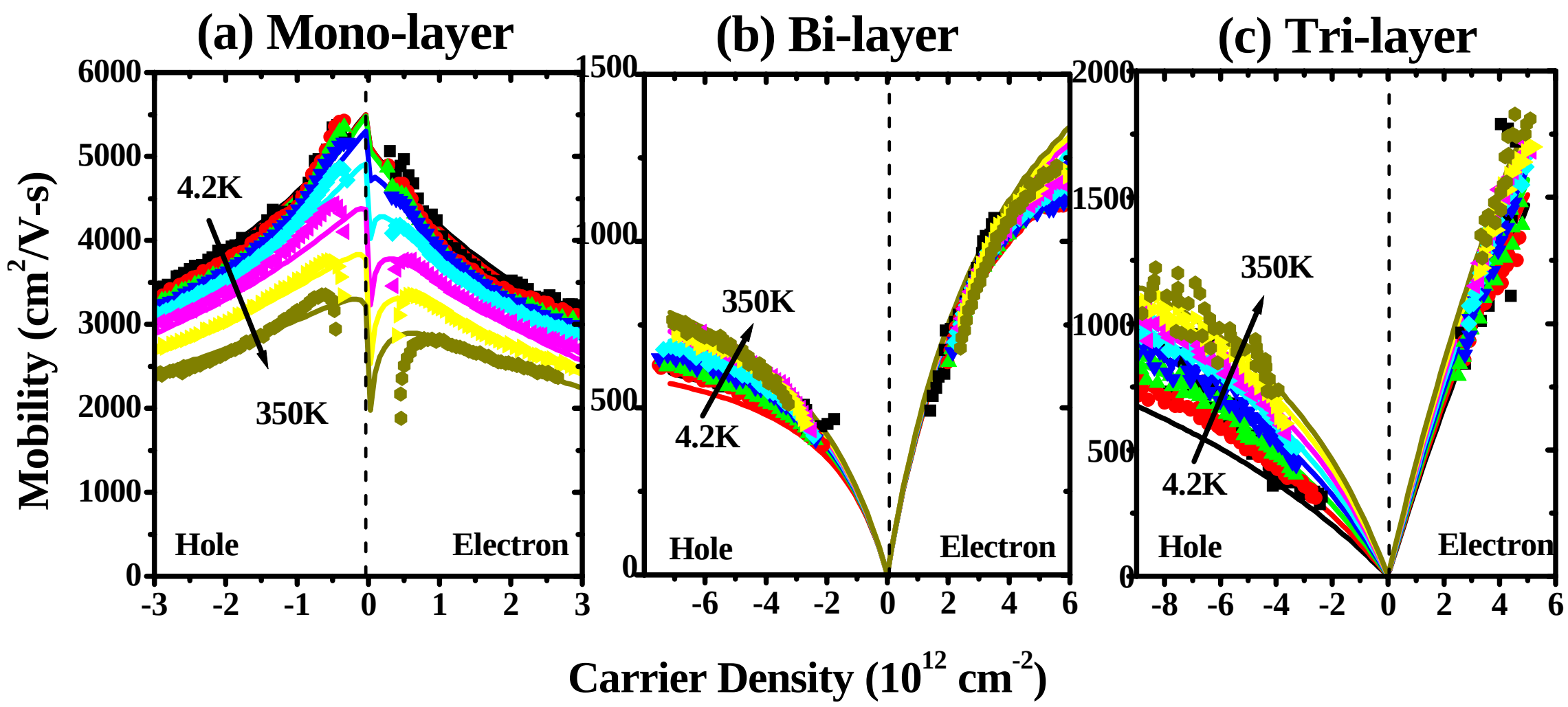

FIG. 1: Hall mobility as a function of carrier density at temperatures from 4.2K to 350K in (a) mono-layer graphene, (b) bi-layer graphene and (c) tri-layer graphene. The symbols are the measured data, the lines are fits.

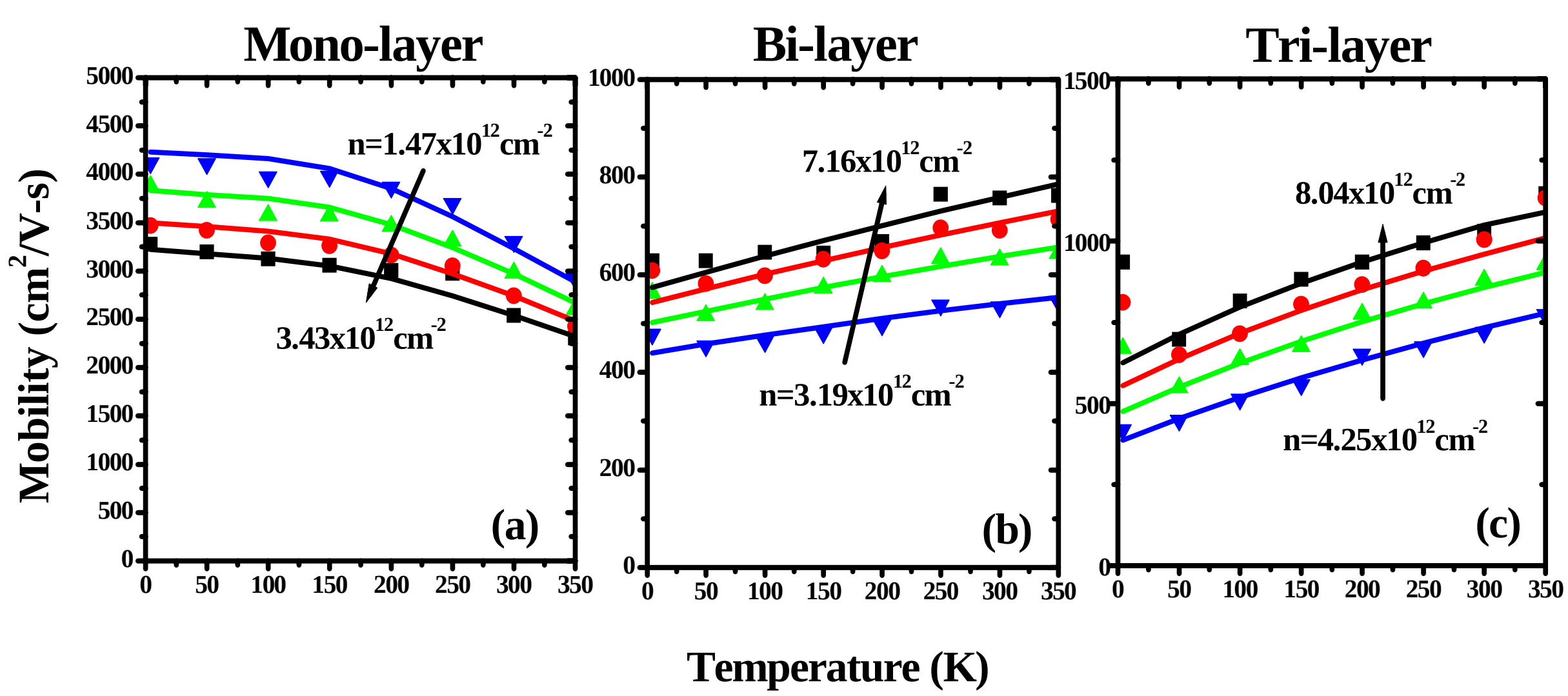

FIG. 2: Hall mobility for holes as a function of temperature at various carrier densities in (a) mono-layer graphene, (b) bi-layer graphene and (c) tri-layer graphene. The symbols are the measured data, the lines are fits.

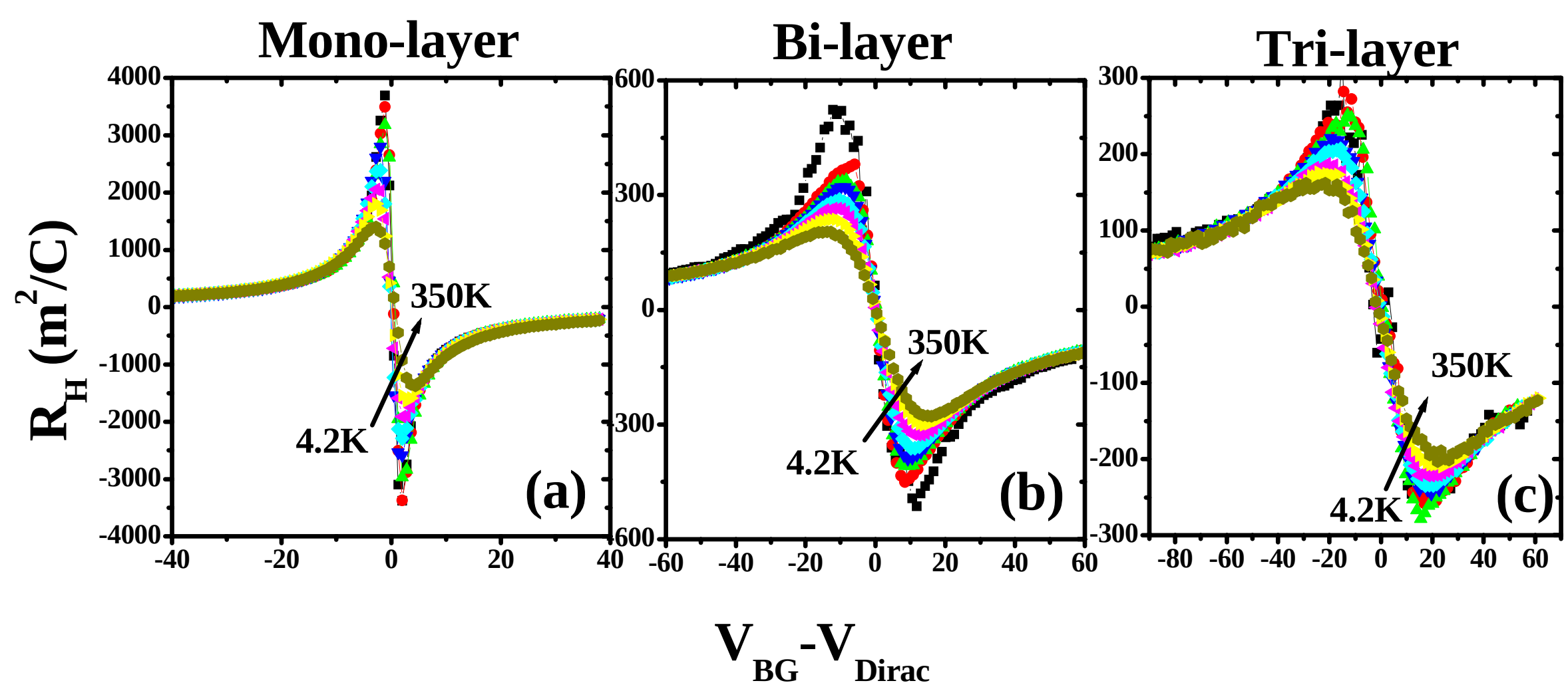

FIG. 3: Hall coefficient as a function of $V_{BG}$-$V_{Dirac}$ in (a) mono-layer graphene, (b) bi-layer graphene and (c) tri-layer graphene.

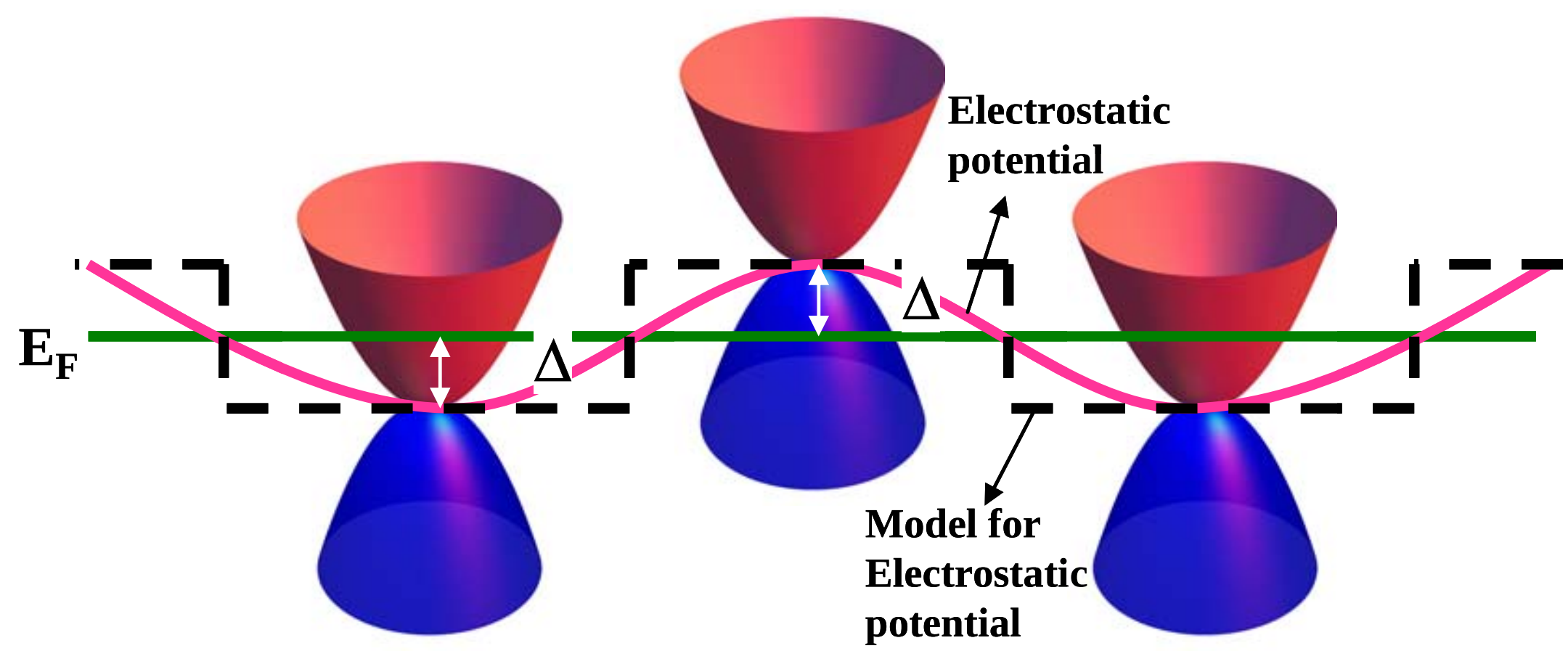

FIG. 4: Illustration of the spatial inhomogeneity of the electrostatic potential and the model used in the analysis of the potential variation for mono-layer, bi-layer and tri-layer graphene.

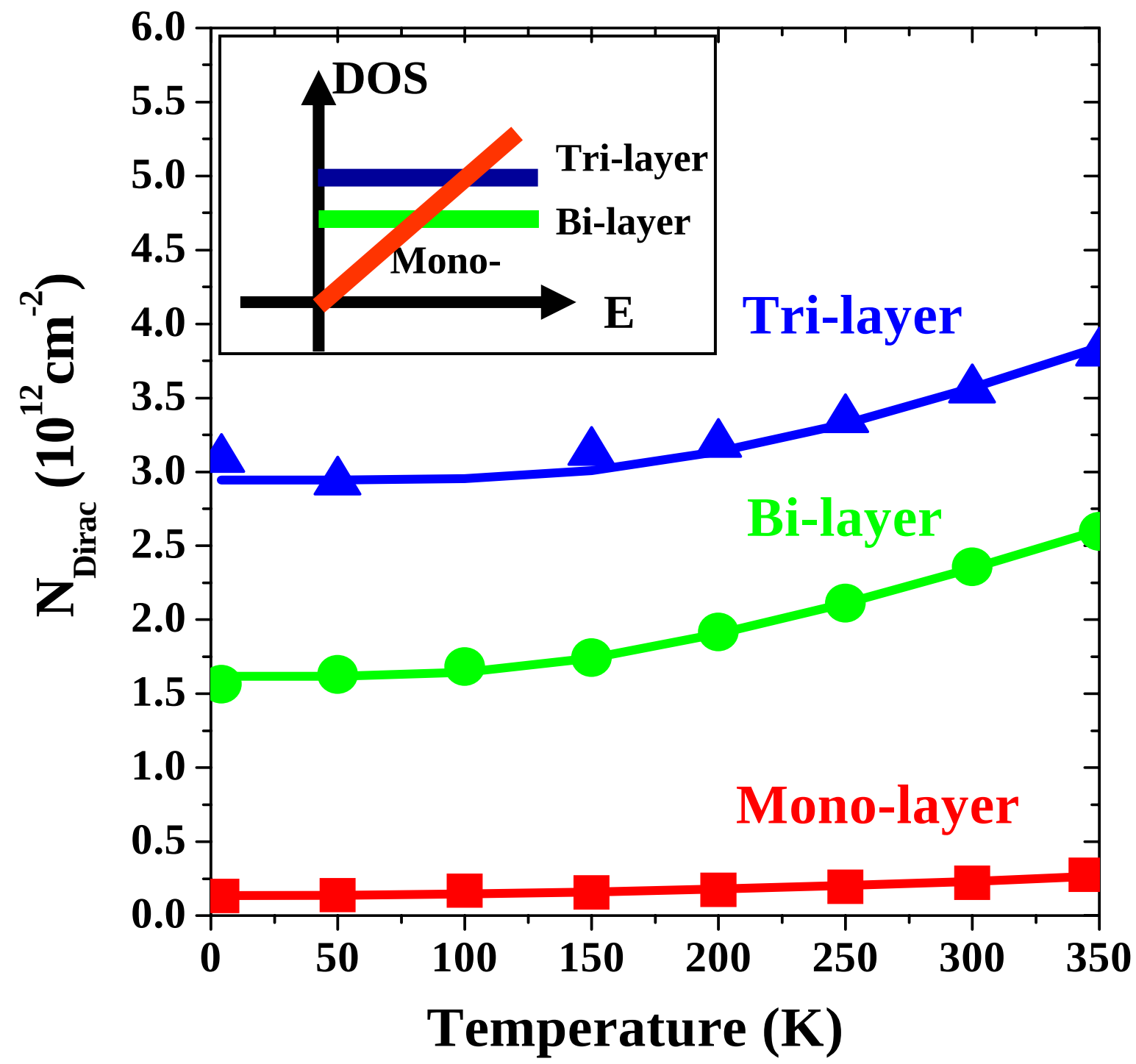

FIG. 5: Carrier density at the Dirac point extracted from the Hall coefficient as a function of temperature in single-, bi- and tri-layer graphene. The symbols are the measured data, the lines are fits using equation (12) for mono-layer graphene and equation (13) for bi-layer and tri-layer graphene.The inset illustrates the density-of-states for single-, bi- and tri-layer graphene.

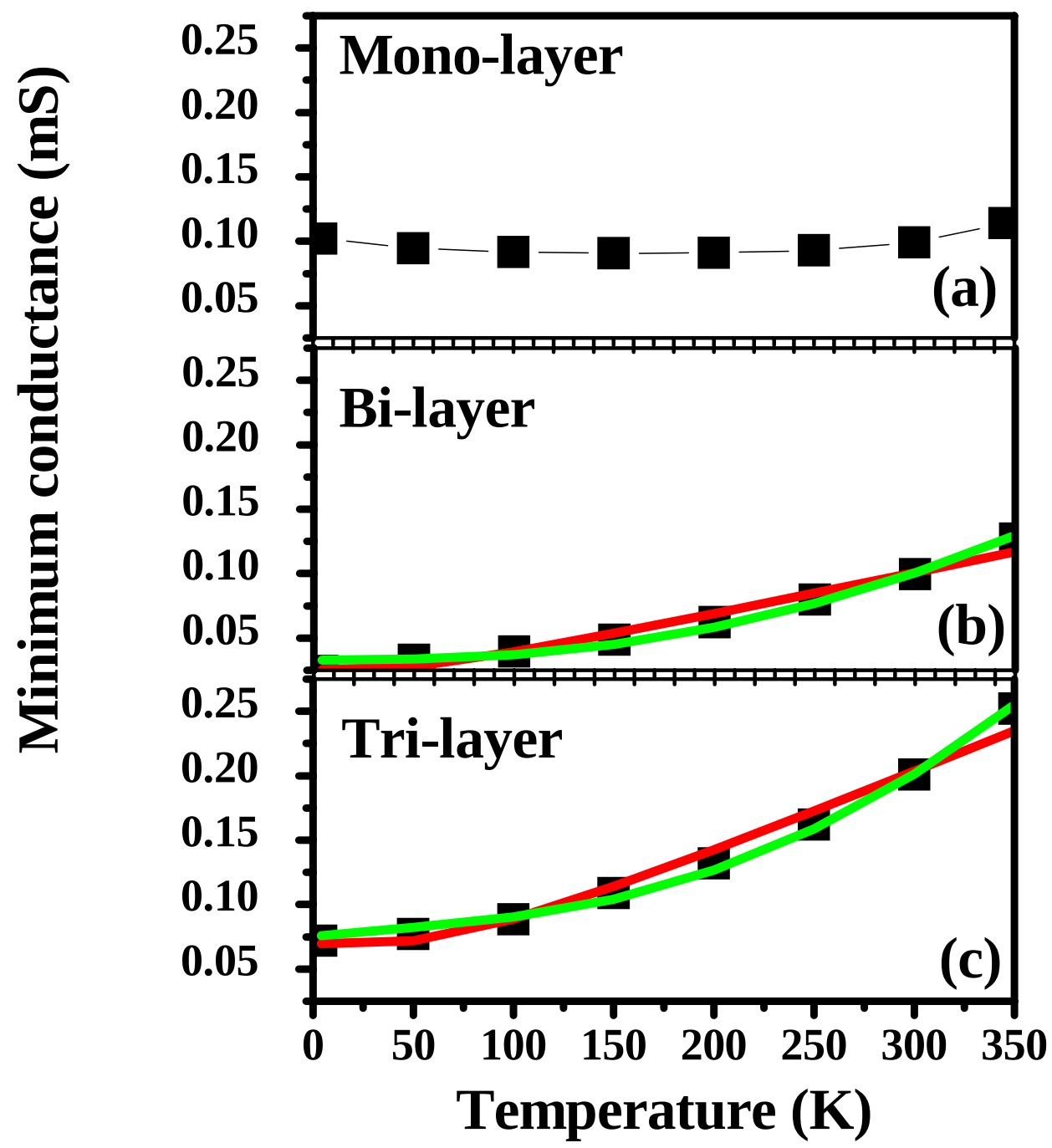

FIG. 6: Minimum conductance as a function of temperature for (a) mono-layer, (b) bi-layer and (c) tri-layer graphene. The symbols are the measured data. In (b) and (c), the red lines are fits that consider only the temperature dependence of the carrier density $n_{Dirac}(T)$ using equation (13), while the green line fits takes into account the temperature dependence of both carrier density and mobility $n_{Dirac}(T)\mu(T)$ using equation (14).

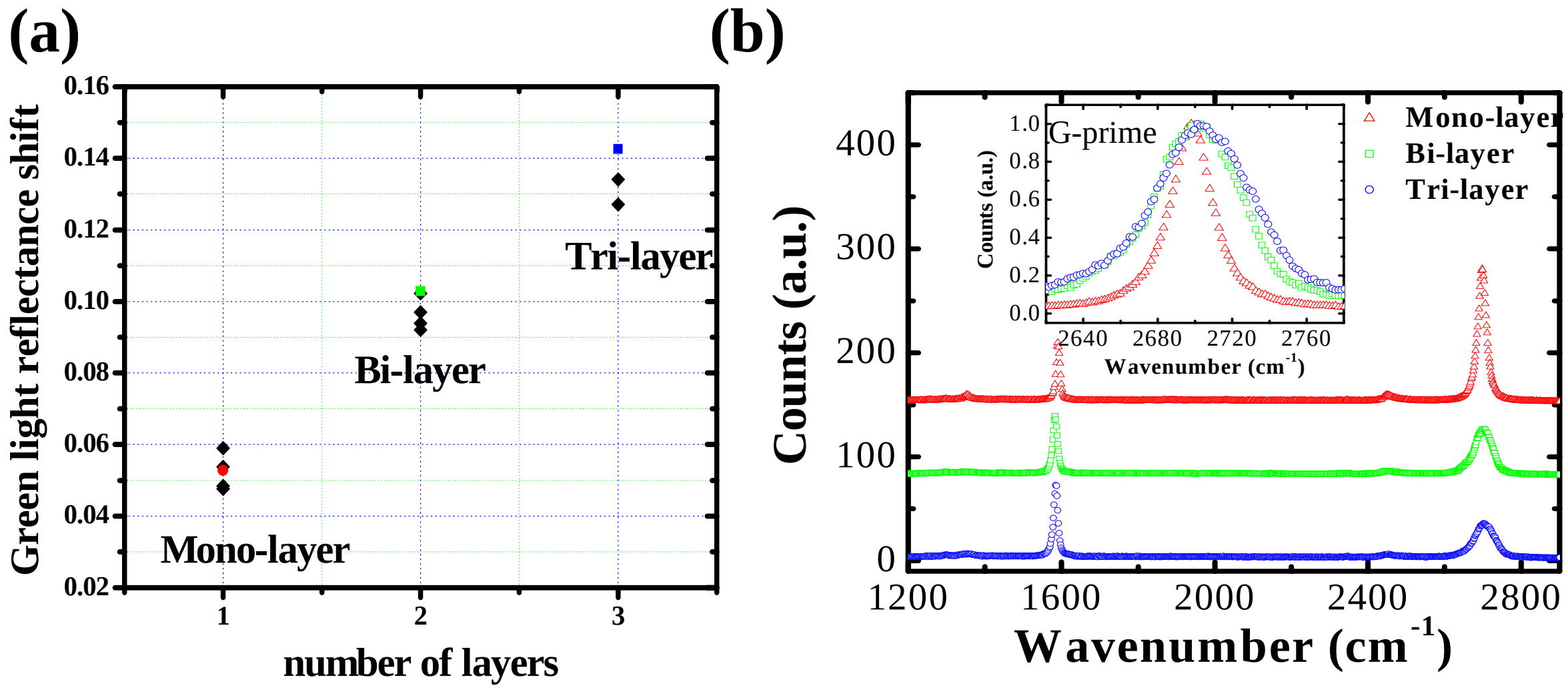

FIG. 7: (a) Green light reflectance shift, (b) Raman spectrum of mono-layer, bi-layer, tri-layer graphenes. The red, green and blue dots in (a) are the flakes measured with Raman as shown in (b). The Raman spectra are offset for clarity. The inset in (b) shows the zoomed-in spectrum of the G-prime band, scaled to the maximum intensity value.

# REFERENCE:

[1] S. A.Wolf, D. D. Awschalom, R. A. Buhrman, J. M. Daughton, S.V. Molnár et al. Spintronics: A Spin-Based Electronics Vision for the Future. Science **294.** 1488 (2001)

[2] D. A. Allwood et.al. Magnetic Domain-Wall Logic, Science **309,** 1688 (2005).

[3] P. Avouris, Z. Chen and V. Perebeinos, Carbon-based electronics. Nature nanotechnology **2,** 605 (2007).

[4] K. S. Novoselov et al. Electric field effect in atomically thin carbon films. Science **306,** 666 (2004).

[5] M. Koshino and T. Ando, Orbital diamagnetism in multilayer graphenes: Systematic study with the effective mass approximation, Physical Review B **76,** 085425 (2007)

[6] E.V. Castro, et al. Biased Bilayer Graphene: Semiconductor with a Gap Tunable by the Electric Field Effect, Phys. Rev. Lett. **99,** 216802 (2007)

[7] T. Ohta, A. Bostwick, T. Seyller, K. Horn, E. Rotenberg, Controlling the Electronic Structure of Bilayer. Science **313,** 951 (2006)

[8] Y. Zhang et al. Direct observation of a widely tunable bandgap in bilayer graphene, Nature **459,** 820 (2009)

[9] M. F. Craciun, S. Russo, M. Yamamoto, J. B. Oostinga, A. F. Morpurgo and S. Tarucha, Trilayer graphene is a semimetal with a gate-tunable band overlap. Nature nanotechnology **4,** 383 (2009).

[10] K. I. Bolotin, K. J. Sikes, Z. Jiang, D. M. Klimac, G. Fudenberg, J. Hone, P. Kim and H. L. Stormer, Ultrahigh electron mobility in suspended graphene. Solid State Communications **146,** 351, (2008)

[11] J. Chen, C. Jang, S. Xiao, M. Ishigami and M. S. Fuhrer. Intrinsic and extrinsic performance limits of graphene devices on $SiO_2$. Nature Nanotechnology **3,** 206, (2008)

[12] E.H. Hwang, S. Adam and S. Das Sarma, Carrier Transport in Two-Dimensional Graphene Layers. Physical Review Letters **98,** 186806 (2007).

[13] E.H. Hwang and S. Das Sarma, Acoustic phonon scattering limited carrier mobility in 2D extrinsic graphene. Physical Review B **77,** 115449 (2008)

[14] V.V. Cheianov and V.I. Fal'ko, Friedel Oscillations, Impurity Scattering and Temperature Dependence of Resistivity in Graphene. Physical Review Letters **97,** 226801 (2006)

[15] Y.-M. Lin, K. A. Jenkins, A. Valdes-Garcia, J. P. Small, D. B. Farmer and P. Avouris, Operation of Graphene Transistors at Gigahertz Frequencies. Nano Letters **9,** 422 (2009)

[16] A. K. Geim and K. S. Novoselov, The rise of graphene. Nature Materials **6,** 183 (2007).

[17] Q. Shao, G. Liu, D. Teweldebrhan and A. A. Balandin, High-temperature quenching of electrical resistance in graphene interconnects. Applied Physics Letters **92,** 202108 (2008)

[18] P. Blake, et al. Making graphene visible. Applied Physics Letters **91,** 063124 (2007)

[19] Y.-W. Tan, Y. Zhang, H.L. Stormer and P. Kim, Temperature dependent electron transport in graphene. Eur. Phys. J. Special Topics **148,** 15 (2007)

[20] K. I. Bolotin, K. J. Sikes, J. Hone, H. L. Stormer and P. Kim, Temperature dependent transport in suspended graphene. Physical Review Letters **101,** 096802 (2008)

[21] T. Ando, Screening Effect and Impurity Scattering in Monolayer Graphene. Journal of Physical Society of Japan **75,** 074716 (2006).

[22] E.H. Hwang and S. Das Sarma, Screening induced temperature dependent transport in 2D graphene. Physical Review B. **79,** 165404 (2009)

[23] A. Akturk and N. Goldsman, Unusually strong temperature dependence of graphene electron mobility. International Conference on Simulation of Semiconductor Processes and Devices (SISPAD), p.173, (2008)

[24] S. Fratini and F. Guinea, Substrate limited electron dynamics in graphene. Physical Review B **77,** 195415 (2008)

[25] T. Stauber, N. M. R. Peres and F. Guinea, Electronic transport in graphene: A semiclassical approach including midgap states. Physical Review B **76,** 205423, (2007)

[26] S.V. Morozov, K.S. Novoselov, M.I. Katsnelson, F. Schedin, D.C. Elias, J.A. Jaszczak and A.K. Geim, Giant intrinsic carrier mobilities in graphene and its bilayer. Physical Review Letters **100,** 016602, (2008)

[27] S. Rotkin, V. Perebeinos, A. Petrov, Ph. Avouris, Nano Lett. **9,** 1850 (2009)

[28] M. V. Fischetti, D. A. Neumayer and E. Cartier, A. Effective electron mobility in Si inversion layers in metal-oxide-semiconductor systems with high-k insulator: The role of remote phonon scattering.

J. Appl. Phys. **90,** 4587 (2001).
[29] D. Ferry, *Transport in Nanostructure* (Cambridge University Press, Cambridge, 2009), Ch.2.
[30] L. Pietronero, S. Stra¨ssler, H. R. Zeller and M. J. Rice, Electrical conductivity of a graphite layer. Physical Review B **22,** 904 (1980).
[31] K. Sugihara, K. Kawamura and T. Tsuzuku, Temperature dependence of the average mobility in graphite. J. Phys. Soc. Jpn. **47,** 1210 (1979)
[32] Y.-M. Lin and P. Avouris, Strong suppression of electrical noise in bilayer graphene nano devices. Nano Lett. **8,** 2119 (2008)
[33] A. Deshpande, W. Bao, F. Miao, C. N. Lau and B. J. LeRoy, Spatially resolved spectroscopy of monolayer graphene on $SiO_2$, Physical Review B **79,** 205411 (2009)
[34] J. Martin et al. Observation of electron–hole puddles in graphene using a scanning single-electron transistor. Nature Physics **4,** 144 (2008)
[35] J. Meyer et al. The structure of suspended graphene sheets. Nature **446,** 60 (2007)
[36] S. V. Morozov, K. S. Novoselov, M. I. Katsnelson, F. Schedin, L. A. Ponomarenko, D. Jiang, and A. K. Geim, Strong suppression of weak localization in graphene. Physical Review Letters **97,** 016801 (2006).
[37] S. Adam. E. H. Hwang, V. M. Galitski and S. Das Sarma. A Self-consistent theory for graphene transport. in *Proceedings of the National Academy of Sciences of the United States of America*, 2007, **104,** p.18392
[38] R. S. Muller and T. Kamins, *Device electronics for integrated circuits, 2nd Edition* (John Wiley & Sons, New York, 1986), Ch.1.
[39] S. Datta, *Quantum Transport: Atom to Transistor* (Cambridge University Press, Cambridge, 2005), Ch.7.
[40] A. Das et al. Monitoring dopants by Raman scattering in an electrochemically top-gated graphene transistor, Nature Nanotechnology. **3**, 210 (2008).